\documentstyle[12pt]{article}
\newcommand{\be}{\begin{equation}}
\newcommand{\ee}{\end{equation}}
\newcommand{\bea}{\begin{eqnarray}}
\newcommand{\eea}{\end{eqnarray}}
\begin{document}
\pagestyle{plain}
\setcounter{page}{1}
\centerline{\bf STRUCTURE AND DYNAMICS OF}
\centerline{\bf MODULATED TRAVELING WAVES IN CELLULAR FLAMES}
\vspace{.75cm}
\normalsize
\centerline{ A. Bayliss, B. J. Matkowsky  and H. Riecke}
\centerline{ {\em Department of Engineering Sciences and Applied Mathematics}}
\centerline{{\em Northwestern University, Evanston, IL 60208}}
\vspace{.50cm}

\small
\underline{\bf ABSTRACT.}
We describe the formation and evolution
of spatial and temporal patterns in
cylindrical premixed flames.  We consider
the cellular regime, $Le < 1$, where the Lewis
number $Le$ is the ratio of thermal
to mass diffusivity of a deficient
component of the combustible mixture.
A transition from stationary, axisymmetric flames to
stationary cellular flames is predicted analytically
if $Le$ is decreased below a critical value. We present the
results of numerical computations to show that as $Le$ is further
decreased, with all other parameters fixed, traveling
waves (TWs) along the flame front arise via
an infinite-period bifurcation which
breaks the reflection symmetry of the cellular array.  Upon further
decreasing $Le$ we find the development
of different kinds of periodically modulated traveling waves (MTWs)
as well as a branch of quasiperiodically modulated traveling waves
(QPMTWs).  These transitions are accompanied by the development
of different spatial and temporal symmetries including
period doublings and period halvings in appropriate coordinate
systems.  We also observe the
apparently chaotic temporal behavior of a disordered
cellular pattern involving creation and annihilation of cells. We analytically
describe the stability of the TW solution near its onset using suitable
phase-amplitude
equations. Within this framework one of the $MTW$'s can be identified
as a localized wave traveling through an underlying stationary, spatially
periodic
structure.

\vspace{.200in}

\normalsize

\noindent
\underline{\bf{1. INTRODUCTION.  }}
It has been long known
that premixed gaseous combustible mixtures do not necessarily burn in a
uniform manner.  In particular, under certain conditions
flame fronts can exhibit a cellular structure, characterized
by periodic arrays of crests along the flame front pointing
in the direction of the combustion products.  The pointed crests
are connected by smooth troughs that are convex
toward the fresh fuel mixture.  The temperature is higher at the
troughs, which therefore appear brighter, and lower
at the crests, which therefore appear darker \cite{mark}.
In many instances the transition from laminar to turbulent
combustion occurs as a smooth flame breaks up into cells,
which then undergo transitions leading to increasingly
complex spatial and temporal behavior.

In \cite{mark} both stationary and propagating cellular
flames were observed
on a slot burner, leading to a linear array of cells,
while rotating cellular flames were observed on a burner of
circular cross section.
A variety of cellular patterns have
been observed for flames stabilized on a circular burner.  In
\cite{gerorder} various patterns of ordered cellular
flames were observed.  For some of the patterns
the cells appeared to be essentially stationary,
although there was a very small
background fluctuation of the intensity,
the origin of which was not clear.
The cells appeared to
be reflection symmetric with respect to both intensity and shape.
In \cite{gherrot} both rotating and modulated rotating cellular
states were observed.  The rotating cellular flames were characterized
by an essentially
uniform rotation rate with no visually detectable
change in the cellular structure
as the cells rotated.  The modulated rotating state exhibited
a modulation in cell size, intensity and
angular velocity.  In both cases the cells
appeared to be asymmetric in
intensity and shape, i.e., the reflection symmetry was broken.

In \cite{gerhop} a different type of dynamical state
was observed, in which one or more
members of a cellular array underwent a rapid motion while
the other members of the array were nearly stationary.  In these
states, termed hopping states, the hopping cells were noticeably
asymmetric while
the nearly stationary
cells appear to be nearly symmetric.  We note that this type of
mode was observed earlier where it was termed a square dance
mode \cite{egrpict}.

In addition to laminar dynamical behavior, chaotic behavior
has also been observed.  In \cite{gerchaos}
four different manifestations of chaotic cellular
flames are characterized.  In particular
disordered states are observed, characterized
by erratic spatial and temporal behavior, exhibiting creation
and annihilation of cells as well as a chaotic variation in
intensity, location and cell size.  Similar behavior has
been observed in \cite{mark}.

One scenario for the development of cellular flames
and the subsequent development of complex spatiotemporal
behavior is that of thermo-diffusive cellular instabilities
\cite{mm,mps,sivashinsky},  which
are obtained from the diffusional
thermal model \cite{msdt} in which the thermal expansion of the gas is
assumed to be weak.
The model accounts for transport and
reaction of heat and one or more components of
the combustible mixture, as well as advection due to the flow field,
which is specified in advance
from a nonreacting flow calculation.
Many of the characteristics of observed cellular flames, including
complex spatiotemporal flame patterns,
can be qualitatively described
as thermo-diffusive instabilities.  Analytical methods have
been successful in describing primary transitions
to cellular flames
(e.g. \cite{gklmgas,mm,mps,om}).
In some cases higher order transitions can be
found by analytical methods as well \cite{omq,om}.
In most instances, however, secondary and higher order transitions
are difficult to find analytically and are more readily
found by numerically continuing the analytically predicted  solution
branches until transitions are found and then following the new
branches.

In this paper we consider cylindrical flames, in particular
adiabatic flames stabilized by a line source
of fuel of strength
$2 \pi \kappa$.  We assume no axial variation and restrict attention
to an axial cross section.  Thus we solve the problem for a fixed
axial slice and employ polar coordinates within this cross section.
We further assume that there is a deficient component
of the combustible
mixture, so that when this component is depleted no further
reaction occurs.  The model then reduces to
two coupled reaction diffusion advection equations.
It is characteristic of combustion
problems that the activation energies are large.  As a result
the reaction terms are important only in a thin layer called
the reaction zone.
In the limit of infinite activation energy the reaction zone
shrinks to a surface called the flame front across which certain
jump conditions hold.  While we consider the finite activation
energy case, for which, strictly speaking, there is no front, we will
use this terminology when appropriate.
Away from the reaction zone the variables change more gradually.
The cellular patterns are most pronounced in the narrow
reaction zone, where the dynamics are very sensitive to the
accuracy of the computation.

The computations build upon and extend
analytical and numerical results previously obtained for this problem.
We first describe the analytical results.  There are
two parameters for this system, $\kappa$ and the Lewis number,
$Le$, the ratio of thermal to mass diffusivity.  In the infinite
activation energy limit this system admits,
for all values of $\kappa$ and $Le$,
a stationary axisymmetric solution
describing a circular flame front
\cite{gklmgas,mps}.
This solution, called the basic solution,
is subject to two different classes of instabilities
depending on whether $Le$ is greater
than or less than 1.

There exists a critical value, $Le_{c1}>1$, such that for
$Le>Le_{c1}>1$, the basic solution is unstable.
The instability arises as two complex
conjugate eigenvalues pass into the right half plane, thus
suggesting that small disturbances evolve to nonstationary
flames, such as axisymmetric pulsating flames, or flames
with traveling or
standing waves along the flame front.  Similar behavior has also
been found in other geometries, e.g. \cite{mm,motravel,mospin,omc}.
The regime $Le$ sufficiently greater than 1, describing, e.g., lean,
heavy hydrocarbon/air mixtures, is often referred to as the
pulsating regime.

Below a second critical value $Le_{c2}<1$
there exists a $\kappa_{c}$ such that for
$\kappa > \kappa_{c}$ the basic solution is also unstable.
This instability occurs by a pair of
real eigenvalues crossing into the right half plane, and
small perturbations evolve to
stationary cellular flames \cite{gklmgas,mps}.  This behavior
has also been found in other geometries \cite{km,mm,omq,sivashinsky}.
The regime $Le$ sufficiently smaller than 1,
describing mixtures where the
deficient component is highly diffusive, for example
lean hydrogen/air mixtures or rich, heavy hydrocarbon/air mixtures, is
referred to as the cellular regime.

In this paper we consider the cellular regime, $Le<1$.
The analysis in \cite{gklmgas,mps} describes
the onset of cellular flames as a diffusional thermal instability.
However the analysis does not describe the subsequent evolution of
these cellular flames into more complex spatial and temporal
patterns and in particular the transition to dynamic behavior
observed in experiments.
These patterns and transitions will be described in this paper
by numerically continuing the stationary cellular solution
branch found in \cite{gklmgas} until new branches are found.
We then follow these new branches to
describe more complex spatiotemporal
behavior.

The analysis in \cite{gklmgas,mps}) identified
the roles of the two fundamental parameters in promoting
diffusional thermal instabilities of cylindrical flames.
For the cellular regime, $Le<1$, the stationary circular flame
front can be destabilized by either increasing $\kappa$ or
decreasing $Le$. In this paper we consider
the behavior as $Le$ is decreased for a fixed value of
$\kappa$.  The behavior of the cellular states as $\kappa$
is increased, with $Le$ fixed, is described in \cite{bmmcell}.

We first describe previous results on
cellular solution branches as $Le$ is reduced.
In \cite{bmmaml} the behavior of stationary cellular flames
was considered.
It was shown that reducing $Le$ led to a progressive
deepening of the cells and more pronounced crests.  Only
stationary solutions were found for the parameter values
used in the computation.  In \cite{bmaml} we considered lower
values of $Le$, and found
nonstationary 4 cell solutions arising via a secondary
transition, i.e., a transition from the branch of
stationary 4 cell flames (S4 branch).  These
solutions described slowly rotating cellular flames. The rotation
occurred at a constant angular velocity and corresponded
to a slowly traveling wave (TW) along the flame front.
It was shown \cite{bmnl}  that the transition
from stationary to rotating behavior occurred as an infinite
period, symmetry (parity)
breaking bifurcation, in which the reflection symmetry
of the cells was destroyed.
That is, while each of the stationary cells was reflection
symmetric about its
centerline, this symmetry was no
longer maintained for the TW solutions.
We refer to this branch as the TW4 branch and will describe
it in more detail below.

The TW4 branch was a pure 4 cell branch, that is, each solution
on this branch was
$2 \pi/4$ periodic in the angular variable $\psi$.  Stable
mixed mode solution branches were
also found and described in \cite{bmnl}.
The spatial behavior
of these mixed mode solutions was that of a 4 cell array
in which at any fixed instant of time the cells differed
in size (e.g., distance between
successive minima of the temperature field)
and shape.  At any fixed spatial location
the temperature and concentration for these
mixed mode cellular flames were quasiperiodic in time.
Two branches of mixed mode solutions
were identified.  The first branch  appeared to
be stable over a small window in $Le$ near the point where the TW4
branch bifurcates from the S4 branch.
The second branch was found for values of $Le$ smaller
than those for which stable TW4 solutions could be found.

In \cite{bmr} we presented preliminary results
identifying the quasiperiodic mixed mode flames
as MTWs,
in which, after a certain time interval
the solution recovers its structure along the front
except for a phase shift \cite{bk}.  Thus the solution
becomes periodic when viewed in an appropriately
chosen rotating coordinate system.
In this paper we extend the results
in \cite{bmr}, identifying additional MTW branches as well
as a new MTW-like branch in which the cells are
quasiperiodically rather than periodically
modulated.

We note that in our computations
the periodic direction is the polar angle $\psi$.  Thus the period
for physically relevant solutions is fixed
($2 \pi$) and is not a parameter
to be varied as in other
problems.  Consequently the number of cells
associated with the MTWs, as well as the
stability of the various solution branches,
is determined by parameters related to the gaseous mixture
(e.g., $\kappa$ and $Le$)
rather than by an assumed period of the solution.
For the parameters considered here,
the MTW and MTW-like branches involve
4 cells rotating about the cylindrical axis.
While results are presented here for a fixed value of
$\kappa$, we believe that similar results, possibly involving
a different number of cells,
would occur for other values of $\kappa$.
There are complex cellular structures and
symmetries associated with each of the branches in addition
to the modulation dynamics.  The cellular structures, including
symmetries, as
well as the dynamics
will be described in detail here.
We also point out the relationship of the calculated MTW
solutions to experimentally observed flames
\cite{gerhop,gherrot}.  While our results are for a simplified
model of combustion, the striking similarities between the
computed modes and those observed in experiments indicates that
the cellular dynamics predicted here does indeed describe
those occuring in real flames.

Specifically,
we identify three distinct MTW branches, each differing from the
others in the nature of the modulation and the symmetries
associated with the modulation of successive cells.
These are referred to as the PMPY
(Pushmi-Pullyu) branch, the BMTW (breathing
MTW) branch and the HPMTW (half-period MTW)
branch, each of which is described below.  In addition,
we identify a quasiperiodically modulated traveling wave
(QPMTW) mode, characterized by a quasiperiodic
modulation involving two apparently independent
frequencies.  Since there is a third frequency
associated with the rotation rate,
this mode has
three apparently
incommensurate frequencies thus describing
a 3-torus.

We now briefly describe the nature of the solutions on
each of the MTW branches.  Solutions along each branch
are composed of 4 cells, which exhibit varied dynamical behavior.
Solutions along the
PMPY branch exhibit a modulation
which is strikingly different from the other modulations that
we observe.
Along the PMPY branch the solution
at any instant of time is composed both of nearly stationary,
nearly reflection symmetric cells and of nonstationary, asymmetric
cells.  Motion of the individual cell proceeds as a localized wave
of asymmetry travels through the underlying
stationary, symmetric cellular array.  This behavior is
analogous to the hopping modes observed experimentally
in \cite{gerhop} and also in
\cite{egrpict} where it was termed the ``square dance mode''.
Along the other two MTW branches the rotation rate about the
cylindrical axis
is a periodic (nearly sinusoidal) modulation of the uniform
rate characteristic of the traveling wave (TW4) branch.  The cells
also undergo a periodic (nearly sinusoidal)
expansion and contraction in both
size and intensity (maximum temperature within the cell).
In a frame moving with the uniform rotation rate the cells
would appear to undergo a breathing motion.  The solution
is characterized by the symmetry that the modulation is the same
for each cell, modulo a constant (in time) phase difference
between any two
cells.  For one of the branches there is no other apparent symmetry.
In view of the breathing motion we refer to this
branch as the BMTW branch (breathing modulated traveling wave).
On the other solution
branch alternate cells undergo a modulation with no
phase difference between them,
so that the solution is periodic over the
interval $0 \leq \psi \leq \pi$.  Since the angular period
has been halved, relative to the BMTW branch,
we refer to this branch as the half-period
modulated traveling wave (HPMTW) branch.
Finally, the QPMTW branch differs from the MTW branches in that
the modulation is quasiperiodic rather than
periodic and the motion appears
to be characterized by three independent frequencies.
The solution exhibits the symmetry that alternate cells
undergo the same modulation with a constant phase
difference between them.  Adjacent cells undergo distinctly
different modulations.
As $Le$ is decreased further the QPMTW branch
as well as the HPMTW branch appear to lose
stability to an attractor which seems to exhibit chaotic temporal
behavior along with a disordered spatial pattern involving
cells undergoing an erratic motion and
creation and annihilation of cells in
an apparently random fashion.  This behavior is characteristic
of weakly turbulent flames.


MTWs have been found previously in other parameter
regimes in combustion as well as in other
physical systems.  In \cite{bmheat} MTWs were found for nonadiabatic
cellular flames in the pulsating
regime.  These modes were shown to be stages
in the development of chaotic behavior for flames
near extinction.  MTWs have been computed for the
Kuramoto-Sivashinsky equation describing
combustion and other areas of application \cite{bk},
and have also been observed in non-reacting flows
e.g., in Taylor-Couette flow experimentally \cite{gs},
and computationally \cite{cm}.

The accuracy of the computed solution is very sensitive to the
resolution of the reaction zone, the location of which is not
known beforehand.  As a result, adaptive procedures are
needed to locate and resolve the reaction zone as the
solution evolves in time.  The problem
of obtaining adequate resolution of the reaction zone is further
complicated by the extremely deep cells which occur as $Le$ is
decreased.  We employ an adaptive pseudo-spectral method
which has been used successfully for a variety of
other problems in combustion (see, e.g., \cite{bgmm,bkm,bmfirst}).
We describe the model in section 2 and
the numerical method in Section 3.
In Section 4 we describe our results in detail.  In section
5 we provide an analytical description of the transition to the TW4
branch within the framework of a resonant mode interaction. In addition,
we employ suitable phase-amplitude equations to discuss its stability
as well as the transition to the PMPY branch.\\[.2in]

\noindent
\underline{\bf{2. MATHEMATICAL MODEL.  }}
We consider the problem of a flame stabilized
by a line source of fuel of strength $2\pi\kappa$.  We assume that
the reaction is limited by a single deficient component and is
governed by one step, irreversible Arrhenius kinetics.  We
denote dimensional quantities by $\tilde{ }$.  The unknowns are the
temperature $ \tilde{T}  $ and the concentration $ \tilde{C}  $ of the
deficient
component.  $\tilde{T}_{u}$ and $\tilde{T}_{b}$ are
the temperatures of the unburned and burned fuel
respectively and $\tilde{C}_{u}$ is the unburned value of $\tilde{C}$.
Other dimensional quantities are the coefficient of thermal diffusivity
$\tilde{\lambda}$, the activation energy
$\tilde{E}$, and the gas constant $\tilde{R}$.
We introduce the nondimensional reduced temperature and nondimensional
concentration by\\[.07cm]
\begin{displaymath}
\Theta=(\tilde{T}-\tilde{T}_{u})/(\tilde{T}_{b}-\tilde{T}_{u}),\ \
C=\tilde{C}/\tilde{C}_{u}.
\end{displaymath}
The spatial and temporal variables are nondimensionalized by
\begin{displaymath}
t=\frac{\tilde{t}\tilde{U}^{2}}{\tilde{\lambda}},\
x_{i}=\frac{\tilde{x_{i}}\tilde{U}}{\tilde{\lambda}},
\end{displaymath}
where $\tilde{U}$ is the planar adiabatic flame speed for the case of
infinite activation energy in which the reaction term is replaced
by a surface delta function.  We employ polar coordinates $(r,\psi)$,
and the nondimensionalized flow velocity due to
the fuel source is ${\bf V}=\frac{\kappa}{r}{\hat{\bf r}}$
where $\hat{\bf r}$ is the unit radial vector.  The equations of the
diffusional thermal model are \cite{mps}
\begin{eqnarray}
\Theta_{t}&=&\Delta \Theta -\frac{\kappa\Theta_{r}}{r}+C\Lambda \, \exp
\left(\frac{Z(\Theta-1)}{\sigma+(1-\sigma)\Theta}\right),    \label{eq:gas} \\
C_{t}&=&\frac{\Delta C}{Le} -\frac{\kappa C_{r}}{r}-C\Lambda \, \exp
\left(\frac{Z(\Theta-1)}{\sigma+(1-\sigma)\Theta}\right).  \nonumber
\end{eqnarray}
Here $\Delta$ is the Laplacian, $\sigma=\tilde{T}_{u}/\tilde{T}_{b},
N= \tilde{E}/(\tilde{R}\tilde{T}_{b})$, $Le$ is the Lewis number, and
$\Lambda=Z^{2}/(2Le)$, where $Z=N(1-\sigma)$ is the
Zeldovich number.
We note that $\Lambda$, which is
referred to as the flame speed eigenvalue, depends on the
nondimensionalization.  The value employed above arises from the use of the
planar, adiabatic flame velocity in the infinite activation energy
limit
$Z \rightarrow \infty$.  A different
nondimensionalization would change the spatial and temporal scales
but would not alter the basic patterns exhibited by the solution.
The boundary conditions are
\begin{eqnarray}
C \rightarrow 1, &&
\Theta \rightarrow 0
\,\  {\rm as} \,\   r \rightarrow 0,  \label{BC1} \\
C \rightarrow 0, &&
\Theta \rightarrow 1
\,\  {\rm as} \,\   r \rightarrow \infty.  \nonumber
\end{eqnarray}
In our computations these boundary conditions are imposed at points
$r_{1}$ and $r_{2}$ far from the reaction zone where combustion occurs.
The computed results were found to be insensitive to changes
in $r_{1}$ and $r_{2}$.

The solution to (\ref{eq:gas})-(\ref{BC1}) has been studied analytically
in the limit $Z \rightarrow \infty$, and $Le - 1 = \rho/Z$
  \cite{gklmgas,mps}.
In this limit the reaction zone shrinks to a surface $r=\Psi (\psi,t)$,
called the flame front.
The following stationary, axisymmetric solution exists for all values
of $\kappa$ and $Le$:
\begin{eqnarray}
\Theta&=&\left\{ \begin{array}{ll}(r/{\kappa})^{\kappa}+O(1/Z),&  r\leq
{\kappa}, \\
1, & r\geq \kappa ,
\end{array}
\right.   \nonumber  \\
C&=&1-\Theta +O(1/Z),   \nonumber  \\
\Psi&=&\kappa, \nonumber
\end{eqnarray}
and is referred to as the basic solution. The dependence of the
solution on $\rho$ enters in the $O(1/Z)$ terms, not written
explicitly here.

This solution becomes unstable if $Le < Le_{c} <1$ and
$\kappa$ is sufficiently large, or if
$Le > Le_{c} >1$.  In the first case a real double eigenvalue
(one corresponding to $\cos$ the other to $\sin$)
crosses from the left half plane into the right half plane
and a transition from the basic
solution to stationary cellular flames occurs \cite{gklmgas,mps}.
Additional transitions have been obtained numerically and are
described in Section 4.\\[.2in]

\noindent
\underline{\bf{3. NUMERICAL METHOD.  }}
We employ an adaptive Chebyshev pseudo-spectral method in $r$ that
we previously developed \cite{bgmm,bmfirst}, together with
a Fourier pseudo-spectral method in $\psi$.
To enhance the resolution of the reaction
zone, we adaptively transform $r$ in order to minimize
a functional monitoring the numerical error.

We first briefly discuss pseudo-spectral methods.  Detailed
descriptions of this class of numerical methods can be found
in \cite{chqz,go}.
We transform the domain for the variable $r$ to the interval $(-1,1)$.
Here we denote the transformed variable by $r$ as well.
The solution, $u$, which generically represents either $\Theta$
or $C$, is then expanded as a sum of basis functions
\begin{equation}   \label{eq:exp}
u \simeq u_{J} \equiv \sum_{j=0}^{J}a_{j} \phi_{j}(r).
\end{equation}
In a Fourier method the basis functions $\phi_{j}$ are trigonometric.
In a Chebyshev method the basis functions $\phi_{j}$ are the
Chebyshev polynomials, i.e.
$\phi_{j}(r)=T_{j}$ where
$T_{j}$ is the $j^{th}$ Chebyshev polynomial,
\begin{displaymath}
T_{j}(r)= \cos (j \cos^{-1}(r)).
\end{displaymath}
The expansion coefficients $a_{j}$ are
obtained from collocation, that is, the function $u_{J}$ is forced to
solve the equations
at a set of $J+1$ collocation points $r_{j}$.
We employ the Gauss-Lobatto points,
\begin{displaymath}
r_{j}=\cos(j \pi /J), \quad 0 \leq j \leq J,
\end{displaymath}
as collocation points.  Thus in a pseudo-spectral method
the unknowns are the solution values at the collocation
points.  The expansion (\ref{eq:exp}) is only used to compute
spatial derivatives.  The Fourier method is similar,
except that trigonometric functions are used
as basis functions.

The major advantage of pseudo-spectral methods over finite difference
methods is enhanced accuracy for a fixed discretization size.
In fact pseudo-spectral methods exhibit infinite order accuracy when
used to approximate infinitely differentiable functions, that is, the
error can be shown to decrease faster than any inverse power of $J$.
These methods are commonly used to approximate problems where
a high degree of accuracy is required, for example in the study
of fluid dynamical instabilities, (see, e.g., \cite{chqz}).
However,
although these methods are highly accurate when used to approximate
functions which exhibit relatively gradual spatial variation, they have
difficulties in approximating functions exhibiting localized regions of
rapid variation, such as the temperature rise and
concentration decay across the narrow
reaction zone.  Severe spatial
oscillations can occur in approximating rapidly varying functions which
are not well resolved \cite{chqz,go}.
These oscillations can affect the computed dynamics
and in certain cases lead to the computation of spurious dynamics
(for example inadequately resolved computations can indicate
chaotic behavior
whereas the exact solution is in fact periodic).

If the location of the reaction zone were known in advance,
the oscillations could be eliminated by introducing
a suitable change of
coordinates so that in the new coordinate system the solution has a more
gradual variation.
However, typically the location of the reaction zone is not
known in advance and is one of the objects of the computation.
In order to realize the benefits (high accuracy) of the pseudo-spectral
method in computing solutions which vary rapidly in localized
regions of space,
we developed an adaptive
pseudo-spectral method.  This method has
proven to be effective in computing the rapidly varying solutions
which occur in combustion problems.
The method is described in detail in \cite{bgmm,bmfirst}.
The description we give here will be brief.

We introduce a family of
coordinate transformations of the form
\begin{equation}
r=q(s,\vec{\alpha})
\end{equation}
where $s$ is the new computational coordinate and $\vec{\alpha}$
represents a parameter vector which is typically of low dimension.
We choose $\vec{\alpha}$  so that in
the new coordinate system the solution exhibits a more gradual variation
and thus is better approximated by a small number of basis functions.
Since the
behavior of the solution changes during the course of the computation,
appropriate values of $\vec{\alpha}$ must be chosen adaptively so as to
adapt to changes in the solution \cite{bgmm,bmfirst}.
The choice of the coordinate transformations, which we discuss
below, is motivated by the asymptotic notion of "stretching" a layer in
the method of matched asymptotic expansions \cite{BO,KC}
for treating singular perturbation problems.

In order to adaptively choose a coordinate transformation in which the
spectral approximation is more accurate we employ error measures which
are computed for each value of $\vec{\alpha}$ until a minimum
is found.  In this paper the error measure is the functional
\begin{equation}   \label{eq:i2}
I_{2}(g)= \left( \int_{-1}^{1}(L^{2}g)^{2}/w(s)ds \right)^{\frac{1}{2}},
\end{equation}
where
\begin{displaymath}
w(s)=\sqrt{1-s^{2}}, \ \ L=w(s)\frac{d}{ds}.
\end{displaymath}
It can be shown that
this functional gives an upper bound on the maximum norm of the error
in approximating a function by its Chebyshev expansion \cite{bgmm}.

A very important feature of the adaptive pseudo-spectral method
is the particular family of mappings.  The unknowns $\Theta$ and
$C$ vary gradually except in and near the reaction zone where rapid
changes occur.  We employ a family of mappings, introduced in
\cite{bt}, in which functions with properties similar to these
are mapped into linear polynomials, which can be
approximated using a relatively small value of $J$.

In order to describe this family of mappings, we let
$I$ denote the interval $-1 \leq r \leq 1$ and consider the
family of functions
\begin{equation}
h(r,\alpha_{1},\alpha_{2}) =
s_{0}+ \tan^{-1}(\alpha_{1}(r-\alpha_{2}))/\lambda.         \label{eq:taninv}
\end{equation}
The parameters $s_{0}$ and $\lambda$ are determined
so that (\ref{eq:taninv}) maps the interval $I$ univalently onto itself,
\begin{displaymath}
s_{0}=\frac{\beta-1}{\beta+1} \ ,
\ \beta= \tan^{-1}(\alpha_{1}(1+\alpha_{2}))/
\tan^{-1}(\alpha_{1}(1-\alpha_{2})),
\end{displaymath}
\begin{displaymath}
\lambda= \tan^{-1}(\alpha_{1}(1-\alpha_{2}))/(1-s_{0}).
\end{displaymath}
When $\alpha_{1}$ is large (the analog of $\epsilon$ small in
singular perturbation problems), (\ref{eq:taninv})
is nearly discontinuous with a region of rapid variation occurring
near $r=\alpha_{2}$, i.e., there is an internal layer at
$r=\alpha_{2}$. We observe that other functions, e.g.,
$\tanh(z)$, can be used
in place of $\tan^{-1}(z)$.

The inverse of (\ref{eq:taninv}),\\[.07cm]
\begin{equation}      \label{eq:tan}
r=q(s,\vec{\alpha})=
\alpha_{2}+ \tan((s-s_{0})\lambda)/\alpha_{1} ,
\end{equation}
describes a two parameter family of mappings of $I$ which are
one to one and onto,
with the property that $h(q(s,\alpha_{1},\alpha_{2}),\alpha_{1},
\alpha_{2})$ is a linear function.  If $\alpha_{1}$ and $\alpha_{2}$
are properly chosen the
temperature and concentration profiles will be sufficiently
similar to (\ref{eq:taninv}) that the composite function
can be represented by a small number of Chebyshev polynomials.
The parameter
$\alpha_{1}$ is a
measure of the rapid rate of change of the function,
while $\alpha_{2}$ is related to the location of the layer, i.e. the
region of rapid variation.  In our method these parameters
are determined adaptively by minimizing the functional (\ref{eq:i2}).
The equations are integrated in time using a first order
splitting method described in \cite{bmnl}.  The timesteps are
kept sufficiently small so that there is no noticeable
effect of temporal integration errors.

In the computation of cellular flames, the location of the
reaction zone depends on $\psi$.  In this case the optimal
values of $\alpha_{1}$ and $\alpha_{2}$ will also depend
on $\psi$.  A two dimensional adaptive pseudo-spectral method
which allows for this dependence has been developed
and applied to problems in gasless condensed phase combustion
in \cite{bkm}.  The resulting coordinate system is nonseparable,
thereby requiring additional computation as described in
\cite{bkm}.  The problem considered here is posed in a
coordinate system in which the Laplacian is separable
and it is more efficient to use
one dimensional coordinate transformations which maintain
separability (i.e. $\alpha_{1}$ and $\alpha_{2}$ independent of $\psi$),
even at the expense of not using values of
$\alpha_{1}$ and $\alpha_{2}$ which are optimal for any particular
value of $\psi$.
We therefore
compute the functional (\ref{eq:i2}) for all angular
points and minimize the average, rather than minimizing for each
value of $\psi$.
We are nevertheless able to obtain a high degree of resolution
of the wrinkled reaction zone due to the effectiveness of
the family of mappings (\ref{eq:tan}) in concentrating
resolution in a fixed region.  As a result, after the
mapping is applied $\Theta$ and $C$ are no longer rapidly varying
in any angular direction although the parameters
$\alpha_{1}$ and $\alpha_{2}$ are generally not optimal
for any particular direction.  \\[.2in]

\noindent
\underline{\bf{4. NUMERICAL RESULTS. }}
In this section we present the results of our numerical simulations of
eqs.(\ref{eq:gas}) at fixed dimensionless activation energy, $N=20$, fixed
temperature ratio of
unburned and burned material, $\sigma =0.615$, and fixed flow rate, $\kappa
=14.8$, while
the Lewis number $Le$ is varied. The timesteps were typically $O(10^{-3})$.  In
all cases
101 Chebyshev collocation points were sufficient to accurately
compute solutions in the radial direction.
We generally used 128 collocation points in the angular
direction, although in some cases we employed 256.  There
was virtually no effect in increasing the number of collocation
points in either independent variable.  In the radial direction the
computational domain was taken as $r_{1} \leq r \leq r_{2}$ with
$r_{1}=1, r_{2}=41$.  We found virtually no effect in changing these
values.
The computations
presented here were obtained at the NCSA and the NERSC.

An overview of the different solution branches
is given in fig.1a where we plot the maximum norm of the
difference between the computed cellular solution and
the (unstable) axisymmetric solution as
a function of $Le$ for the different
solution branches that we have found.
We generally concentrate on determining the nature of
the transitions that occur and on the properties of solutions along
different solution branches rather than on determining the precise
numerical values of the control parameter
$Le$ at which the transitions occur.
We note that we solve
the initial value problem, marching
forward in time until a steady state is achieved,
so that in general we only compute
stable solutions.  However unstable solutions can sometimes
be computed by modifying the computer program
so as to impose certain symmetries in the angular
direction.  For example, unstable axisymmetric solutions can
be computed by allowing no variation in $\psi$.  Unstable
4 mode solutions can be computed by restricting the computation
to the interval $0 \leq \psi \leq 2 \pi/4$ and imposing periodicity.
Unstable reflection symmetric solutions can be computed by
restricting to the interval
$0 \leq \psi \leq 2 \pi/8$ and imposing reflection
symmetry.  We have not computed unstable solutions which are not
stabilized by the imposition of specific symmetries.

In the regime considered here we find 3 branches of stationary
solutions,
S4, S5 and S8, which have, 4, 5 and 8 cells
respectively.  The cells associated with solutions along these branches
are reflection symmetric.
Only the S4 branch
bifurcates stably from the basic axisymmetric state;
the S5 and S8 branches are unstable at onset.
The S5 branch becomes stable,
however, for
smaller $Le$. This behavior is not unexpected since in  general only
the state closest to the minimum of the neutral curve is stable to sideband
instabilities
at onset, whereas the other states become stable at the
Eckhaus stability (stability to sideband
disturbances) limit. When decreasing $Le$ the S4 branch
continuously merges with the S8 branch, i.e.
the first harmonic becomes increasingly stronger
whereas the fundamental eventually
goes to zero. Thus, the S4 branch bifurcates
from the S8 branch in a pitchfork bifurcation.
We will not discuss
the behavior
of solutions that arise when the S5 branch becomes unstable.
We also did not follow the S8
branch for values of $Le$ below those
indicated in fig.~1a.

Before reaching the S8 branch,
the S4 branch becomes unstable to perturbations which break
the reflection symmetry (parity-breaking bifurcation) \cite{bmnl}.
To calculate solutions on the S4 branch beyond that
point we therefore restrict the solutions to be reflection symmetric.
At the parity-breaking bifurcation point a branch of traveling 4 cells,
TW4, emanates from the
S4 branch.
These waves are similar to the rotating cells observed in \cite{gherrot}
which were also found to be asymmetric while
the nonrotating cells were symmetric. In
the computations the rotation rate and the degree of asymmetry
\cite{bmnl}
go to 0 continuously as the bifurcation point is approached, indicating that
this
is an infinite period bifurcation.
Strikingly, the TW4 solutions
appear to be unstable at onset, although the
bifurcation corresponds to a supercritical pitchfork bifurcation.
Perturbed solutions evolve either to
a solution on
the S5 branch
or, in a small window of $Le$, to a solution on a MTW branch
which we refer to as the
``Pushmi-Pullyu'' (PMPY) branch, as discussed below.
The relevant instability has the character
of a sideband instability and can
be suppressed by restricting the domain of computation
to $0 \le \psi \le 2\pi /4$ and imposing periodicity.
In the full domain
the TW4 solutions
eventually become stable for smaller $Le$
($Le \leq 0.37$). The instability close to
onset may explain why the rotation rates of the waves observed in
the experiments \cite{gherrot} do not approach 0 continuously as
the transition point is approached.

When $Le$ is decreased further, the TW4 solutions
become unstable again and a branch of modulated waves arises
in what appears to be a supercritical Hopf bifurcation.
Due to the visual appearance of a
breathing of the cells we call this 2 frequency state a BMTW.
For even smaller values of $Le$ two
additional, different branches of modulated waves arise;
the QPMTW, which arises from the BMTW branch
and appears to be a quasiperiodically modulated wave
with three frequencies, and the HPMTW branch which bifurcates from
the (unstable) TW4 branch.
These branches are shown in detail
in fig.~1b.
In the following we discuss the properties of the modulated
waves in detail.

We first consider the PMPY branch. Fig.~2 shows a contour plot of $\Theta$ at
fixed
$t$ for $Le=0.42$. The contours have
been chosen to emphasize contours of $\Theta$ corresponding
to values near 0.98
where the cellular structure of the solution is most pronounced.
Clearly, none of the 4 cells are exactly alike; two of the cells are nearly
reflection symmetric, characteristic
of the behavior along the branch S4, while two of the cells
are not symmetric, characteristic of the behavior along
the branch TW4.
The dynamics of this solution is shown in fig.~3
where $\Theta(r_{*},\psi,t)$ is shown
as a function of $\psi$ for increasing values of $t$.
Here and in the following $r_{*}$ is chosen so that the circle $r=r_{*}$ is in
the reaction zone.  As the forward crest (temperature minimum)
of a given cell (cell 3 in the figure if we number
the cells from the left)
is jerked clockwise (to the left in fig.~3), the cell expands,
becomes asymmetric and
moves clockwise. This, in turn, pulls the cell behind it (cell 4)
so that it too expands, becomes asymmetric and moves clockwise.
At the same time cell 3 pushes the cell ahead of it,
compressing it and causing it to become nearly symmetric and nearly
stationary. This process then repeats with cells 1 and 2 etc. in turn and can
be viewed
as a {\it localized} wave of asymmetry propagating counter-clockwise through
the stationary
symmetric cellular array. This dynamics suggest that the solution is best
described as a
``Pushmi-Pullyu'' \cite{dolit} solution.
Its behavior is analogous to that of the hopping modes observed
experimentally in \cite{egrpict,gerhop}.
Similar patterns have been observed in directional
solidification \cite{fsl91}, in directional viscous fingering
\cite{coud} and in Taylor vortex flow \cite{wm92}.

Alternatively, it can be seen from fig.~3 that the dynamics of the PMPY
solution
 can also be characterized as that of a MTW, since after a discrete time
interval $\tau$
the pattern reemerges, though
shifted in space,
\be
\Theta (\psi - c \tau ,t+\tau)=\Theta (\psi,t). \label{e:qpsym}
\ee
 Thus, in a coordinate system rotating uniformly with speed $c$, $\tilde{\psi}
= \psi - c t$,
the solution would be periodic. In addition, the PMPY solutions
possess the more restrictive
symmetry
\be
\Theta (\tilde{\psi} + 2\pi/4 ,t+\tau/4)=\Theta (\tilde{\psi},t),
\label{e:pomsym}
\ee
which connects adjacent cells. It implies that all 4 cells perform the
same dynamics although shifted in time.
In the context of a discrete or cellular $k$-mode system with
periodicity $2 \pi$, the symmetry
\be
\Theta (\tilde{\psi} + 2\pi/k ,t+\tau/k)=\Theta (\tilde{\psi},t),
\label{e:pomsymk}
\ee
referred to as ``ponies on a merry-go-round'' (POM),
is often observed \cite{agmp,bk}.

The BMTW solutions exhibit the same symmetry (\ref{e:pomsym}) as the
PMPY solutions.
The form of modulation, however, is different.
This is shown in fig.~4 which illustrates the temporal evolution of
$\Theta(r_{*},\psi,t)$.
In order to understand why the TW4 branch does not persist stably
for lower values of $Le$,
we note that the general effect of decreasing
$Le$ is to increase the size of the cells, where by the
size of the cells we refer to the elevated high temperature
region between the two successive
minima.  Decreasing $Le$ corresponds to
increasing the diffusivity
of the reactant relative to the diffusivity of heat.  As a result
the reactant diffuses into a region of higher temperature where
a higher degree of burning occurs.  This in turn generates
more heat which then diffuses, thus raising the temperature
over a larger region, which effectively increases
the size of the cells.  For the values of $Le$ considered,
the radial location of the reaction zone does not
change significantly. Therefore the cells
must fit into the circumference $2 \pi r_{*}$.  As $Le$ is
decreased, progressively larger cells are forced to fit
into the fixed circumference.
Beyond a certain point a more stable configuration
is that illustrated in fig.~4 where some cells have contracted
while their neighbors have expanded.
Analysis of the spectrum of the computed solutions indicates that the
BMTW branch appears to arise due to a sideband instability
whereby TW4 solutions are modulated by a mode 1 modulation.

The following dynamical behavior is observed.
Similar to the PMPY solutions
the cells undergo an overall rotation with
a modulation of cell size and intensity.
In contrast to the PMPY solutions, the rotation rate never gets close
to 0 and all the cells are asymmetric at all times.
Thus the asymmetry is global rather than localized
as for the PMPY solutions.
In order to illustrate
the nature of the modulation, we consider the maxima of
$\Theta(r_{*},\psi,t)$.
We find that for each fixed $t$, $\Theta$ attains exactly
4 maxima
for $0 \leq \psi < 2 \pi$
which we denote by $\hat{\Theta_{i}} (i=1, \ldots 4)$.
Clearly each such maximum
is identified with an individual cell and may be thought
of as a measure of the intensity of the cell.
We note that we have accounted for the possibility that any
particular maximum may well lie between two angular collocation
points.  We do so by accumulating all of the maxima for which
$\psi$ is fixed over a time interval and then taking the largest
of these maxima.
In fig.~5a the temporal evolution of
$\hat{\Theta_{i}}$ is plotted for 2 adjacent cells (i.e.,
$i=1,2$).  The figure clearly shows both that the modulation
is periodic and is essentially the same for the two cells
except for a constant phase shift.  Similar properties
hold for the other two cells as well.  In addition, the
phase shift is the same between all pairs
of adjacent cells. In fig.~5b the angular
location of each maximum is shown as a function of $t$.
Clearly, in addition to their breathing motion the cells
undergo a periodic oscillation in
position and rotation rate.

Breathing type MTW solutions,
somewhat analogous to the BMTW solutions found here, have also been
found for nonadiabatic, pulsating ($Le>1$) flames near extinction
\cite{bmheat}.  These MTWs also appeared to arise
via a bifurcation from a TW flame, specifically a
7 cell TW which was simultaneously destabilized by the
subharmonic modes 3 and 4.
However the MTWs found in \cite{bmheat} exhibited
a different symmetry from the POM symmetry (\ref{e:pomsymk}).  They
exhibited the symmetry referred to as
``jumping ponies on a
merry-go-round'' (JPOM) \cite{bmheat},
\begin{equation}    \label{eq:jpom}
\Theta (\tilde{\psi} +  l \times 2 \pi/k ,t+\tau/k)=\Theta (\tilde{\psi},t).
\end{equation}
Whereas the POM symmetry relates each cell to the cell adjacent
to it, this symmetry relates cells according to the sequence
$1, l+1, 2l+1, ... ({\rm mod} \; k)$, effectively skipping the $l-1$
cells between successive members of the sequence.
Here $k=7$ corresponds to the total number of cells. Thus, for example,
if at a certain time cell 1 attains its maximum temperature then
$1/7th$ of a period later in time cell $1+l$ would attain its maximum.
In an experiment a temperature maximum
is observed as a spot of maximum brightness.
Thus, with $l \neq 1$ the bright spot would be seen as
jumping over the $l-1$ cells between cells $1$ and $1+l$.  The POM
symmetry is a special case of (\ref{eq:jpom}) with
$l=1$.  We note that the
general symmetry (\ref{eq:jpom}) has been discussed in
\cite{r} in the context of rotating fluids.
In \cite{bmheat} transitions to
JPOM MTWs with $l=2$,
with $l=4$ and with $l=1$ (corresponding
to the POM symmetry (\ref{e:pomsymk})) were found.
Thus MTWs exist in two different parameter regimes
within the same model of combustion, however the mechanism
of instability  as well as the symmetries of the resulting
MTW, is different in the two cases.  In both
cases there are additional transitions leading to chaos,
but the nature of the transitions is different for the two regimes.


When $Le$ is reduced to about $Le =0.21$ the BMTW branch
becomes unstable and a transition to the
QPMTW branch occurs, as the symmetry
(\ref{e:pomsym}) is broken.  Near the transition point the
symmetry breaking is most pronounced in the phase
differences between pairs of adjacent cells.  This is shown in
fig.~6a where we plot $\Theta(r_{*},\psi,t)$ for $Le$=0.205.
It can be clearly seen that
cells 1 and 2 (numbered from the left) have a small phase
difference between them (in fact they appear to be nearly
in phase).  Cells 3 and 4 behave
similarly.  In contrast, the
phase difference between cells 2 and 3 and
between cells 4 and 1 is considerably larger.  As $Le$ is further
reduced from the transition point, the differences in the
amplitude of the modulation between
the odd and even cells becomes more pronounced.  This can be seen
in fig.~6b where we plot $\Theta(r_{*},\psi,t)$ for $Le$=0.180. In addition,
each cell undergoes a modulation in size, similar to the dynamics
of the BMTW solutions. However,
adjacent cells now undergo different modulations and the 4 cell
array splits into two pairs.  Within each pair the modulations
are nearly in phase while there is a large phase difference
between the two pairs.  This behavior has the character of a spatial
period doubling. We remark that in order to facilitate
a comparison between the BMTW, QPMTW and HPMTW
branches, both the total temporal duration and the
time difference between successive curves, is the same for
figs.~6a and 6b as well as for the analogous figs.~4 and 8.

While figs.~6a and 6b give the visual appearance of a periodic
modulation, upon closer examination we find that the
amplitude of the modulation is quasiperiodic and thus
the waves are quasiperiodically modulated.
This can be seen in figs.~7 where we examine the temporal evolution
of the maxima $\hat{\Theta{_i}}$. In figs.7a,b the evolution
of adjacent and alternate maxima, respectively,
is illustrated.  For each cell the amplitude modulation
is quasiperiodic, with a high frequency oscillation modulated
by a low frequency envelope.  Alternate cells have the same envelope,
and their modulations differ only by a constant phase
shift with respect to the envelope.  Adjacent
cells have distinctly different envelopes.  In fig.~7c we plot
the power spectral density for the cell with the larger amplitude
modulation (i.e. cell 2 in fig.~6b when ordered from the left).
In addition to the main peak at $\omega_{1} \simeq 0.263$ we
find a low frequency peak at $\omega_{2} \simeq 0.016$ which
corresponds to the slow modulation of the envelope.  These two
frequencies do not appear to be related to each other by a simple
rational relation.
Since this spectrum is calculated from the
maximal temperature of one particular cell,
it represents the dynamics in a frame rotating with that cell.
In the laboratory frame the rotation rate
will add a third frequency to the spectrum.
The location of the maxima $\Theta_i$ as a function of time is shown in
fig.~7d.  We
note that there is a modulation in position (and thus
in velocity) and that the modulations are distinctly
different for adjacent cells.

Following the QPMTW branch further by decreasing $Le$,
we find a transition to solutions
exhibiting seemingly random creation and annihilation of cells
through phase slips.  For values of $Le$ near
the transition point, the chaotic
behavior is characterized by intervals where cells are
created and annihilated without undergoing any organized
motion, alternating, in a seemingly
random fashion, with intervals in which there are
4 cells which undergo a roughly organized motion
similar to that of QPMTW solutions.  This general type
of behavior is qualitatively similar to
the intermittently ordered states observed
in \cite{gerchaos}.  We illustrate this behavior in figs.~10a-b
where we plot
$\Theta(r_{*},\psi,t)$ for $Le$=0.165.  The figures are
sequential and contiguous in $t$, the data is presented in
2 separate figures for clarity.
Note that although a transition to chaos can generally be
expected from a branch of
quasiperiodic solutions with
three independent frequencies \cite{bpv},
here the transition may not be via
the quasiperiodic route
as a result of symmetries.
This is also
supported by the fact that
phase slips are very important
in the chaotic dynamics so that
the qualitative features of the chaotic
attractor strongly differ from those of the quasiperiodic attractor.

We also find a narrow window in $Le$ in
which a fourth branch of stable modulated waves exists.
The behavior of solutions along this branch is illustrated
in fig.~8 where $\Theta (r_{*},\psi,t)$ is plotted for $Le=0.175$.
Again, the shape of the cells is modulated in time.
In contrast to the BMTW and QPMTW solutions,
however, alternate cells are {\it in phase},
i.e. the solution has the discrete translation
symmetry
\begin{displaymath}
\Theta(\psi+2\pi/2,t)=\Theta(\psi,t),
\end{displaymath}
which does not involve any phase shifts. Thus
the spatial period is only half the system size, as
compared to the BMTW solutions
which only have the period $2\pi$.
We therefore call this new branch a half-period MTW (HPMTW).

Our results indicate that the HPMTW branch
bifurcates from the TW4 branch. Since the
TW4 solutions have the spatial symmetry
\begin{equation}    \label{eq:p4sym}
\Theta(\psi+2\pi/4,t)=\Theta(\psi,t),
\end{equation}
the bifurcation corresponds to a pure period doubling in space.
For comparison, the
analogous transition from the TW4 branch to the BMTW branch
constitutes a period quadrupling.
The modulation of the
temperature maxima is strictly periodic as can be seen
from fig.~9a where the temperature maxima for two adjacent
cells are plotted.  The angular locations of the 4 maxima
are plotted in fig.~9b as a function of time.  Note that the
rotation direction is reversed from the previous cases.  This
is an effect of the initial conditions.  For each
rotating solution, a solution rotating in the opposite direction
can easily be obtained by simply reflecting the angular coordinate.
We also observe from
fig.~9b that the oscillation in peak location is exactly
in phase for alternate cells.

The HPMTW branch can be obtained by following the (unstable)
TW4 branch, imposing periodicity over the interval
$0 \leq \psi \leq \pi$.
At onset the HPMTW branch is unstable when considered
over the full angular
interval $0 \leq \psi \leq 2 \pi$.
This is expected since it bifurcates from the
TW4 branch which is itself unstable for this range of $Le$.  From the
 computation in the restricted domain we have observed that the subharmonic
mode (mode 2) appears to grow continuously from 0 as $Le$ is
reduced thus indicating that the HPMTW branch bifurcates from the
TW4 branch.  In the full domain the HPMTW branch becomes stable
for lower values of $Le$. Both in the restricted and the full
domain the HPMTW branch loses stability to
an apparently chaotic attractor for $Le$ approximately 0.165. Thus, its range
of stability
extends beyond that of the QPMTW branch and at $Le=0.170$ the HPMTW
branch coexists stably with an
apparently chaotic solution arising from an instability of the QPMTW
solutions for $Le$ sufficiently small.

We note that the two successive bifurcations of modulated
waves from the TW4 branch (to the BMTW
and HPMTW branches, respectively)
suggests the possibility
of a third bifurcation, to a branch which, extending our notation
might be called a quarter period MTW, in which each cell
is modulated but all cells are exactly in phase and
satisfy the symmetry (\ref{eq:p4sym}).
In fact the modulated wave reported in \cite{gherrot} appears
to satisfy an analogous symmetry (the
number of cells was different).  An investigation of the
TW4 branch for smaller values of $Le$ has failed to find
such a branch although it is quite possible that such a branch
exists for other parameter values.

\vspace{1cm}

\noindent
\underline{\bf{5. ANALYSIS.  }}
In this section we describe two approaches which yield insight into the
connection between the S4, S8 and TW4
branches as well as between the TW4 and PMPY branches, respectively. The
merging of
the S4 branch with the S8 branch suggests considering the resonant
interaction of a mode with amplitude $A$ and wave number $q$ and a mode
with amplitude  $B$ and
wavenumber $2q$ corresponding to S4 and S8, respectively.
In the vicinity of a codimension-2 point in parameter space at
which both these modes simultaneously destabilize
the axisymmetric state
the dynamics of the system restricted to $2\pi/4$
periodic solutions can be described by two complex amplitude equations
\begin{eqnarray}
\partial_TA&=&\mu_1 A+c_1A^{*}B+a_1A|A|^2+b_1A|B|^2, \label{e:q2q} \\
\partial_TB&=&\mu_2 B+c_2A^2+b_2B|A|^2+a_2B|B|^2, \nonumber
\end{eqnarray}
where $^*$ denotes complex conjugate, and $T$ is a slow time variable.
The coefficients $a_{i}$, $b_{i}$, and $c_{i}$ are
$O(1)$, and
$\mu_{1}$ and $\mu_{2}$ are the control parameters.
These equations, which have been studied extensively, e.g.
\cite{arm,dang,pj}, exhibit a parity-breaking transition from a
stationary $q$ mode branch to a TW branch before the $q$ mode branch
merges with the $2q$ mode branch, just as in the numerical
computations of cellular flames considered here.
Though these simulations are not close to such a
codimension 2 point, calculations in directional
solidification \cite{rr} and in Taylor vortex flow \cite{rp}
showed that the regime of existence of the TW branch can extend far
beyond this special point in parameter space.
Thus the amplitude equations (\ref{e:q2q}) for the $q:2q$ mode interaction
appear
to describe transitions from S4 to TW4 and to S8, respectively.
The amplitude equations also describe MTW solutions, but these
are different from the MTW solutions described in this paper.
This is due to the fact that eqs.(\ref{e:q2q}) are restricted
to a description of the behavior of four identical cells, while
in our MTW solutions PMPY, BMTW and HPMTW as well as in our QPMTW
solution, the four cells do not behave identically.

To understand the stability behavior of solutions
on the TW4 branch as well as the relation of the TW4 branch to
the PMPY branch,
it is useful to take a different approach. In the vicinity of a
parity-breaking instability the dynamics can be described by two coupled real
equations for the amplitude $A$ of the small
antisymmetric part of the
solution, which measures the extent to which the reflection symmetry
is broken, and the local phase $\phi$ of the solution, which gives the
instantaneous position of each cell \cite{cgg89}. In the context of the
present computation the symmetric solution S4 corresponds to $A=0$ and
the asymmetric TW4-solution to $A=A_0=const$. In order to describe the
PMPY-solution
one would have to allow the asymmetric amplitude to depend on the spatial
coordinate. This is systematically possible only in the limit that the
spatial variations occur on a long length scale.
We therefore imagine introducing the radius $R$ of the circular flame front as
an additional
parameter which can be changed by changing the flow rate $\kappa$. In the limit
$R \rightarrow \infty$ the introduction of a slow angular coordinate
$X = \epsilon \psi$ and a slow time variable $T= \epsilon t$, $\epsilon \ll 1$,
 on which $A$ and $\phi$ are allowed to depend,
is justified and one can find evolution equations for
$A$ and $\phi$. One may hope, that these equations then also describe the
system with
finite $R$ in a qualitative way. Strikingly, this is indeed the case as is
shown
below.

To derive the evolution equations for the amplitude $A$ of the asymmetric
part and the local
phase $\phi$ we expand the physical variable, say $u$,  as
\begin{displaymath}
u = u_{S}(\phi) +\epsilon A(X,T) u_{A}(\phi) + h.o.t.,
\end{displaymath}
and use symmetry and scaling arguments to obtain the evolution
equations \cite{ccf92,cgg89}
\begin{eqnarray}
\partial_T \phi&=&A, \label{e:aphi} \\
\partial_T A&=&\lambda_1 \partial_X \phi A+b_2 \partial_{XX} \phi \nonumber \\
& &+ \epsilon \left\{ (\lambda_0 +\lambda_2 (\partial_X \phi)^2)A+d_2
\partial_{XX} A-A^3+gA\partial_XA+
b_{21} \partial_X \phi \partial_{XX} \phi \right\}. \nonumber
\end{eqnarray}
Note that the equations explicitly
contain $\epsilon$ which signifies that, strictly
speaking, two separate slow time scales may be called for. Instead, we
consider the reconstituted equations and keep track of the relative order of
the
terms as signified by $\epsilon$.
Extended traveling wave solutions,
which we associate with TW4 solutions, are given by
\be
A^2 = \lambda_0 \equiv A_0^2, \ \  \phi_0 = \omega T
\ee
with $\omega = A_0$. In general, $\phi$ could also have a linear dependence on
$X$, but
it can be absorbed into the coefficients $\lambda_1$, $\lambda_2$ and $b_{21}$.
 To determine the linear stability
of this solution we write it as
\be
A = A_0 + \delta A_1 e^{ipx+\sigma t}, \ \ \phi = \phi_0 + \delta \phi_1
e^{ipx+\sigma t}.
\ee
Inserting into (\ref{e:aphi}) and linearizing in $\delta \ll 1$
yields the dispersion relation
\be
\sigma^2+\epsilon \sigma \left( 2 A_0^2 + d_2 p^2 - ipgA_0 \right) + b_2 p^2
-ipA_0 \lambda_1 =0.
\ee
Using the Hurwitz criterion, the TW solutions are stable iff
\be
d_2 >0
\ee
and
\bea
H(p,A_0) &\equiv&  b_2 d_2^2 p^4  + p^2 d_2 A_0^2 \left(4 b_2  +\lambda_1 g
\right)
\nonumber \\
& &+ \; 2  \lambda_1 g A_0^4  + 4 b_2 A_0^4 -
\lambda_1^2 A_0^2 \epsilon^{-2} >0. \label{e:hurwitz}
\eea
Thus, stability with respect to perturbations
with large wave numbers $p$  (short waves) requires
$b_2 >0$, which is also a condition for
the stationary structure described by $A=0$
to be Eckhaus-stable. If $b_{2} < 0$,
a fourth derivative of $\phi$ must be included in
(\ref{e:aphi}),
as in (\ref{e:aphieck}) below,
in order to obtain a well posed problem.
Strikingly, if $\epsilon$ is small,
traveling wave solutions are unstable to long
wavelength perturbations due
to the term involving
$\lambda_1$ \cite{ccf92}.  That is, for
$\epsilon$ and $p$ small, (\ref{e:hurwitz})
cannot be satisfied.  Thus, in an infinite system,
 which allows $p \rightarrow 0$, i.e., allows
very long-wave perturbations, the traveling waves
are in general {\it unstable at onset} if $\lambda_1 \ne 0$. Since
$\lambda_1$
determines the slope of the neutral stability
curve corresponding to the
secondary bifurcation, this stability
behavior is similar to that of stationary structures which
arise in a primary bifurcation;
there, only the
structure with the critical wavenumber (at the minimum of the
neutral curve, where the analog of $\lambda_{1}$ is 0)
is stable at onset with respect to the Eckhaus instability. States with other
wave numbers
can become stable further above threshold.
By analogy, we might expect that
the TW branch could also gain stability at larger
amplitudes, even if $\lambda_1 \ne 0$. This can be seen by assuming
$\lambda_1$ to be small,
\be
\lambda_1 = \epsilon \Lambda_1.
\ee
Since $H$ is strictly increasing with $p^2$,
the TW solutions
become stable in an infinite system when $H(p=0)=0$, i.e. for
\be
A_0^2 \ge \frac{\Lambda_1^2}{4 b_2}.
\ee

The TW branch $A^{2}=\lambda_{0}$
arises through a supercritical pitchfork
bifurcation. In a finite system
with periodic boundary conditions the bifurcation
is also a pitchfork and we therefore
expect the TW solutions
to be stable even at onset, since only a single
real eigenvalue passes through 0.
The eigenvalue corresponding to the translation mode
is identically zero and, since the spectrum is discrete, the remaining
eigenvalues are in the left half plane,
bounded away from the imaginary axis.
This result can be recovered from
(\ref{e:hurwitz}) by limiting $p$ to $p \ge p_c \equiv 2 \pi/L$ where $L$ is
the
length of the system. Then for small
$\epsilon$ the TW branch
is stable in two windows given by
\bea
A < \frac{1}{4} \sqrt \frac{\Lambda_1^2}{b_2}\left(1-\sqrt{1-\frac{8b_2
d_2}{\Lambda_1^2} \frac{4 \pi^2}{L^2}}\right),\\
A > \frac{1}{4} \sqrt \frac{\Lambda_1^2}{b_2}\left(1+\sqrt{1-\frac{8b_2
d_2}{\Lambda_1^2} \frac{4 \pi^2}{L^2}}\right).
\eea
The TW branch becomes stable for all amplitudes,
i.e., the window of instability disappears, if
\be
L^2 < \frac{32 b_2 d_2 \pi^2}{\Lambda_1^2}.
\ee
Based on (20) we might therefore expect that the TW solution is
stable at onset.

In contrast to this argument, in the numerical
computation of (\ref{eq:gas}) the TW4 branch
seems to be unstable as close to onset as could be investigated.
In addition, the stationary S4 branch
appears to be stable all the way to the parity-breaking bifurcation,
beyond which, presumably due to the Eckhaus instability,
perturbations evolve to the S5 branch.
We therefore consider the case that the parity-breaking
instability and the Eckhaus
instability ($b_2 =0$) occur extremely close to
one another. Such a situation occurred
in Taylor vortex flow \cite{rp}. Thus we extend
equations (\ref{e:aphi}) to
the case $b_2 = \epsilon^2 \beta$ and allow $\beta$ to have
either sign depending on which bifurcation occurs first.
This leads to
\begin{eqnarray}
\partial_T \phi&=&A, \label{e:aphieck} \\
\partial_T A&=&\lambda_1 \partial_X \phi A \nonumber \\
& &+\epsilon \left\{ (\lambda_0 +\lambda_2 \ (\partial_{X}\phi)^2)A+d_2
\partial_X^2 A-A^3+gA\partial_XA+
b_{21} \partial_X \phi \partial_X^2 \phi \right\}  \nonumber \\
& &+\epsilon^2 \left\{ \beta \partial_X^2 \phi - b_4 \partial_X^4 \phi
\right\}. \nonumber
\end{eqnarray}
Note that we assume $b_{4}>0$ for well posedness.
Taking $\lambda_1 =\epsilon^2 \Lambda_1$ leads to
\bea
H&= &\left( 4 (\beta + b_4 p^2) +2 \Lambda_1 g\right) A_0^4     \nonumber \\
& &+ \left(d_2 p^2 ( 4 \beta + 4 b_4 p^2 + \Lambda_1 g) - \Lambda_1^2 \right)
A_0^2 +
d_2^2 p^4 \left( \beta + b_4 p^2 \right).
\eea
Thus, for small amplitudes the TW solutions
are unstable to long-wavelength perturbations
$p^2 < - b_4/\beta$, as in the
stationary structure. However, for larger amplitudes
the unstable TW solutions become stable if
\be
\Lambda_1 g > - 2 \beta > 0.
\label{e:gstab}
\ee
Thus, with $\beta < 0$ the above scenario describes a TW branch
which is unstable at onset but stabilizes at some
distance from onset,
as found in the numerical
computation of (\ref{eq:gas}).

The coupled phase-amplitude equations (\ref{e:aphieck}) also give
insight into the PMPY solution discussed in section 4.
The PMPY solutions
can be described as a localized wave of asymmetry traveling through
the underlying symmetric array of
stationary cells. For the case $b_2 >0$, for which the stationary
cells are Eckhaus-stable,
it has been shown that (\ref{e:aphi})
supports stable solutions of that kind
\cite{ccf92,rp}. Their existence can be seen easily in
the special case
$g=b_{21} = \lambda_2 =0$ \cite{rp}. After introducing $q=\partial_X \phi$ and
going into
a frame moving with velocity $v$, (\ref{e:aphi}) can be solved
for $q$, as
\be
q=-A/v + q_\infty. \label{e:qA}
\ee
For stationary solutions this leads to an `equation
of motion for a particle with mass $d_2$',
\be
d_2 \partial_X^2 A + \left( v - \frac{b_2}{v} \right) \partial_X A =
-\partial_A U(A)
\label{e:particle}
\ee
in the `potential'
\be
U(A) = \frac{1}{2} (\epsilon \lambda_0 + \lambda_1 q_\infty ) A^2 -
\frac{1}{3} \frac{\lambda_1}{v} A^3 -\frac{\epsilon}{4} A^4.
\ee
Localized solutions correspond to homoclinic orbits of
(\ref{e:particle}) which
start at the maximum of the potential at $A=0$ for $X = - \infty$, reach a
turning point at $A_{max}$ and return to $A=0$ for $X = \infty$.
Such orbits exist if the `friction' $v - \frac{b_2}{v}$ vanishes and if
$\lambda_0 \ge \lambda_c \equiv - \lambda_1/\epsilon \, ( 2\lambda_1/(9\epsilon
b_2)+
q_\infty)$. The latter condition
ensures that the value of the potential at its second maximum,
which occurs at nonzero $A$,
is positive and therefore above the value at the turning
point $A_{max}$, $U(A_{max})=0$.

Since the numerical calculations in section 4
strongly suggest that the stationary structure
becomes Eckhaus-unstable very close to the parity-breaking bifurcation, the
analysis
of (\ref{e:aphi}) is not sufficient. Applying a similar
analysis to the
extended eqs.(\ref{e:aphieck}) leads to an integral condition
for the determination of $v$, given by
\bea
\int_{-\infty}^\infty \,\partial_X A \left(
\left ({\frac {\epsilon\,b_{{21}}}{v^{2}}}+
\epsilon\,g\right )A \partial_X A + \left (v-\frac{b_2}{v}-\frac{
\epsilon\,b_{21} q_\infty}{v}\right )\partial_X A
+\frac{\epsilon^2 b_4}{v}\,\partial_X^3 A  \right)\,dX \nonumber \\
 = 0,
\eea
which replaces the condition
$v - b_{2}/v=0$.
This condition expresses the statement that there is no change
in the total `energy' along the
homoclinic orbit. Instead of treating this condition together
with the equation of motion by perturbation
methods \cite{ccf92}, we solve (\ref{e:aphieck}) numerically, which,
in addition to showing the existence of such localized solutions,
will also show their stability, at least for the given parameter
values.

Localized waves behave like a symmetric stationary state ($A=0$) for
$X \rightarrow \pm \infty$, i.e. for $q \rightarrow q_\infty$,
and like an asymmetric traveling wave ($A^2=\lambda_0$) in
a localized region. From (\ref{e:qA}) with $A^2 = \lambda_0$ the wave number in
the
asymmetric region is $q_{as} \neq q_{\infty}$.
We would like to determine parameter values for use in the
numerical solution of
(\ref{e:aphieck}), to exhibit stable localized
waves. Thus we would like to determine conditions on the parameter values
to insure that the symmetric stationary state is stable for
$X \rightarrow \pm \infty$ ($q \rightarrow q_\infty$), and that
the asymmetric traveling wave is stable for $X$ in the
asymmetric region ($q=q_{as}$). It is simpler to determine
conditions that insure that the symmetric stationary and
the asymmetric traveling wave solutions, each defined on the entire
$X$ interval, are simultaneously stable. These conditions then
provide an estimate for the parameter values to be used in
(\ref{e:aphieck}).
For the stability of the symmetric stationary state we
require
\be
b_2 + b_{21} q_\infty > 0, \ \  d_2 > 0, \ \ \ \lambda_1 q_\infty  + \epsilon(
\lambda_{0} +\lambda_{2} q_{\infty}^{2}) < 0,
\ee
where the first condition corresponds to the condition $b_{2}>0$ given
above, with the role of $b_2$ played by
$b_2 + b_{21} q_\infty$,
and the third condition follows from the requirement that the growth rate
(coefficient of the term linear in $A$ in (\ref{e:aphieck})) is negative.
For the stability of the asymmetric traveling wave we choose
$g$ to satisfy
\be
\Lambda_1 g > - 2 (b_2 + b_{21} q_{as}) > 0,
\ee
which is the analog of (\ref{e:gstab}),
where now $b_{2}+b_{21}q_{as}$ plays the role of $b_{2}$.
Thus, though the stationary state at the local wave number
$q_{as}$ would be Eckhaus-unstable
since $b_2 + b_{21} q_{as} <0$, the traveling wave can be stable
due to the presence of the
nonlinear gradient term $g A \partial_X A$. Fig. 11a shows the temporal
evolution of
$A$ for such a stable localized wave with $\lambda_0 = -0.1$, $\lambda_1 = -3$,
$\lambda_2 = 0$, $d_2 =1$, $b_2 =0$, $b_{21}=0.1$, $b_4 = 1$ and $g=-0.5$. In
fig.~11b the amplitude and the
wave number are shown for a representative time.

 Unfortunately, the coefficients for (\ref{e:aphi})
are not known for the
system of interest here. Thus, a quantitative comparison is not possible. The
above
calculations show, however, that these equations may very well describe the
behavior
of the cellular flames in the vicinity of the parity-breaking bifurcation.

\vspace{1cm}

\underline{\bf{ACKNOWLEDGMENTS.  }}
This work was supported
by N.S.F. grants MSS 91-02981 and DMS 90-20289 and
D.O.E. grants DE-FG02-87ER25027 and DE-FG02-92ER14303.
We are pleased to acknowledge useful
discussions with M. Golubitsky, M. Gorman and I. Kevrekidis.

\pagebreak
\centerline{Figure Captions}
\vspace{.1in}
\noindent
Figure 1a. Solution branches as a function of $Le$.\\[.3cm]
\noindent
Figure 1b. Solution branches over restricted
region in $Le$.\\[.3cm]
\noindent
Figure 2. Contour plot for $\Theta$ for solution on PMPY branch,
$Le=0.42$.\\[.3cm]
\noindent
Figure 3. $\Theta(r_{*},\psi,t)$ for solution on PMPY branch,
$Le=0.42$, $r_{*}=13.5$.
Time $t$ (angle $\psi$) increases along vertical (horizontal) axis. \\[.3cm]
\noindent
Figure 4. $\Theta(r_{*},\psi,t)$ for solution on BMTW branch,
$Le=0.23$, $r_{*}=10.4$.
Time $t$ (angle $\psi$) increases along vertical (horizontal) axis. \\[.3cm]
\noindent
Figure 5a. Temperature of two adjacent maxima (peaks) vs. $t$,
for solution on BMTW branch, $Le=0.27$, $r_{*}=11.158$.\\[.3cm]
Figure 5b. Angular location of temperature
maxima (peak) vs. $t$, for
solution on BMTW branch, $Le=0.27$, $r_{*}=11.158$.\\[.3cm]
\noindent
Figure 6a. $\Theta(r_{*},\psi,t)$ for QPMTW solution , $Le=0.205$,
$r_{*}=10.2$.
Time $t$ (angle $\psi$) increases along vertical (horizontal) axis. \\[.3cm]
\noindent
Figure 6b. $\Theta(r_{*},\psi,t)$ for QPMTW solution , $Le=0.185$,
$r_{*}=10.0$.
Time $t$ (angle $\psi$) increases along vertical (horizontal) axis. \\[.3cm]
\noindent
Figure 7a. Temperature of two adjacent maxima (peaks) vs. $t$,
for solution on QPMTW branch, $Le=0.185$, $r_{*}=9.842$.\\[.3cm]
Figure 7b. Temperature of two alternate maxima (peaks) vs. $t$,
for solution on QPMTW branch, $Le=0.185$, $r_{*}=9.842$.\\[.3cm]
Figure 7c. PSD corresponding to Fig. 7a.\\[.3cm]
Figure 7d. Angular location of temperature
maxima (peaks) vs. $t$, for
solution on QPMTW branch, $Le=0.185$, $r_{*}=9.842$.\\[.3cm]
Figure 8. $\Theta(r_{*},\psi,t)$ for solution on HPMTW branch,
$Le=0.175$,
$r_{*}=9.8$.
Time $t$ (angle $\psi$) increases along vertical (horizontal) axis. \\[.3cm]
Figure 9a. Temperature of two adjacent maxima (peaks) vs. $t$,
for solution on HPMTW branch, $Le=0.175$, $r_{*}=9.346$.\\[.3cm]
Figure 9b. Angular location of temperature
maxima (peak) vs. $t$, for
solution on HPMTW branch, $Le=0.175$, $r_{*}=9.346$.\\[.3cm]
Figure 10a. $\Theta(r_{*},\psi,t)$ for solution apparently exhibiting
apparently chaotic temporal behavior with a disordered
spatial structure , $Le=0.165$,
$r_{*}=9.6$.
Time $t$ (angle $\psi$) increases along vertical (horizontal) axis.\\[.3cm]
Figure 10b. $\Theta(r_{*},\psi,t)$ for solution apparently exhibiting
apparently chaotic temporal behavior with a disordered
spatial structure , $Le=0.165$,
$r_{*}=9.6$
Time $t$ (angle $\psi$) increases along vertical (horizontal) axis.
Continuation of fig.~10a.\\[.3cm]
Figure 11. \\
a) Temporal evolution of asymmetry amplitude $A$ according to phase-amplitude
equations (\ref{e:aphieck}) for $\lambda_0 = -0.1$, $\lambda_1 = -3$,
$\lambda_2 = 0$,
$d_2 =1$, $b_2 =0$, $b_{21}=0.1$, $b_4 = 1$ and $g=-0.5$.\\
 b) Representative stable localized wave at same parameters as a).

\newpage
\begin{thebibliography}{99}
\bibitem{arm} D. Armbruster, J. Guckenheimer and P. Holmes,
{\it Heteroclinic cycles and modulated travelling waves
   in systems with O(2) symmetry}, Physica D {\bf 29} (1988), 257-282.
\bibitem{agmp} D. G. Aronson, M. Golubitsky and J. Mallet-Paret,
{\em Ponies on a merry-go-round in large arrays of josephson
junctions}, Nonlinearity {\bf 4} (1991), 903-910.
\bibitem{bgmm} A. Bayliss, D. Gottlieb, B. J. Matkowsky, and M.
 Minkoff, {\em An adaptive pseudo-spectral method for reaction diffusion
problems,}
J. Comput. Phys. {\bf 81} (1989), 421-443.
\bibitem{bkm} A. Bayliss, R. Kuske and B. J. Matkowsky, {\em A
two-dimensional adaptive pseudo-spectral method},
J. Comput. Phys. {\bf 91} (1990), 174-196.
\bibitem{bmfirst} A. Bayliss and B. J. Matkowsky, {\em Fronts,
relaxation oscillations, and period doubling in solid fuel
combustion, }J. Comput. Phys. {\bf 71} (1987), 147-168.
\bibitem{bmaml} A. Bayliss and B. J. Matkowsky, {\em Spinning
cellular flames, }Appl. Math. Lett. {\bf 3} (1990), 75-79.
\bibitem{bmminn} A. Bayliss and B. J. Matkowsky, {\em Bifurcation
pattern formation and chaos in combustion, } in ``Dynamical
Issues in Combustion Theory'', P. Fife, A. Linan and F. Williams eds.,
Springer-Verlag (1991) 1-36.
Proceedings on Dynamical Issues in Combustion Theory.
\bibitem{bmger} A. Bayliss and B. J. Matkowsky, {\em Bifurcation
pattern formation and transition to chaos in combustion, } in
``Bifurcation and Chaos: Analysis, Algorithms,
Applications'', R. Seydel, F. Schneider, T. Kupper and H. Troger eds.,
Birkh\"auser, Basel (1991), 36-51.
\bibitem{bmnl} A. Bayliss and B. J. Matkowsky,
{\em Nonlinear dynamics of cellular flames,}
SIAM J. Appl. Math., {\bf 52} (1992), 396-415.
\bibitem{bmheat} A. Bayliss and B. J. Matkowsky,
{\em From traveling waves to chaos in combustion, }
to appear, SIAM J. Appl. Math..
\bibitem{bmmaml} A. Bayliss, B. J. Matkowsky and M. Minkoff, {\em
Adaptive pseudo-spectral computation of cellular flames stabilized
by a point source,} Appl. Math. Lett. {\bf 1} (1987), 19-24.
\bibitem{bmmcell} A. Bayliss, B. J. Matkowsky and M. Minkoff, {\em
Cascading cellular flames,} SIAM J. Appl. Math. {\bf 49} (1989), 1421-
1432.
\bibitem{bmmfrance} A. Bayliss, B. J. Matkowsky and M. Minkoff, {\em Numerical
computation of bifurcation phenomena and pattern formation in combustion,}
in ``Numerical Combustion'', A. Dervieux and B. Larrouturou eds., Lecture Notes
   in Physics 351 , Springer-Verlag, Heidelberg (1989), 187-198.
\bibitem{bmr} A. Bayliss, B. J. Matkowsky and H. Riecke, {\em Modulated
traveling waves in combustion,} in Numerical Methods for PDEs
with Critical Parameters, eds. H. G. Kaper and M. Garbey,
Kluwer, Dordrecht (1992), 137-162.
\bibitem{bt} A. Bayliss and E. Turkel, {\em Mappings and accuracy
for Chebyshev pseudo-spectral approximations, }J. Comput. Phys.
{\bf 101} (1992), 349-359.
\bibitem{BO} C.M. Bender and S.A. Orszag, {\em Advanced
Mathematical Methods for Scientists and Engineers} McGraw
Hill, New York, (1978).
\bibitem{bpv} P. Berge, Y. Pomeau and C. Vidal, {\em Order within
Chaos, }John Wiley \& Sons, N. Y. (1984).
\bibitem{bk} H.S. Brown and I.G. Kevrekidis, {\em A computer
assisted study of modulated travelling waves in the Kuramoto-
Sivashinsky equations, }to appear in Proceedings of the Workshop
of Applications of Pattern Formation, J. Chadam, W. Langford and
B. Wetton eds., Fields Institute for Research in Mathematical
Sciences, (1993).
\bibitem{chqz} C. Canuto, M. Y. Hussaini, A Quarteroni and T. A. Zang,
{\em Spectral Methods
in Fluid Dynamics, } Springer-Verlag, New York, (1987).
\bibitem{ccf92} B. Caroli, C. Caroli and S.Fauve,
{\em On the Phenomenology of Tilted Domains in Lamellar Eutectic Growth},
 J. Phys. {\bf I 2} (1992), 281-290.
\bibitem{cm} K. T. Coughlin and P. S. Marcus, {\em Modulated
waves in Taylor-Couette flow Part 2. Numerical simulation},
J. Fluid Mech. {\bf 234} (1992), 19-46.
\bibitem{cgg89} P. Coullet, R.E. Goldstein and G.H. Gunaratne,
{\em Parity-breaking transitions of modulated patterns in hydrodynamic
systems}, Phys. Rev. Lett. {\bf 63} (1989), 1954-1957.
\bibitem{dang} G. Dangelmayr, {\it Steady-state mode interactions in the
presence of O(2)-symmetry}, Dyn. Stab. Sys. {\bf 1} (1986), 159-185.
\bibitem{egmb} M. el-Hamdi, M. Gorman, J. W. Mapp and J. I.
Blackshear Jr., {\em Stability boundaries of periodic modes of
propagation in burner-stabilized methane-air flames, }Comb. Sci.
and Tech. {\bf 55} (1987), 33-40.
\bibitem{egrpict} M. el-Hamdi, M. Gorman and K. A. Robbins, {\em A picture book
of dynamical modes of flat, laminar premixed flames, }Technical
Report \#2, University of Houston (1990).
\bibitem{fsl91} J.M. Flesselles, A.J. Simon and A. Libchaber,
{\it Dynamics of one-dimensional interfaces: an experimentalist's view},
Adv. Phys. { \bf 40} (1991), 1.
\bibitem{gklmgas} M. Garbey, H. A. Kaper, G. K. Leaf and
B. J. Matkowsky, {\em Linear stability analysis of cylindrical
flames, }Quarterly of Applied Math. {\bf 47} (1989), 691-704.
\bibitem{gerorder} M. Gorman, M. el-Hamdi and K. A. Robbins,
{\em Experimental observation of ordered states of
cellular flames, }preprint.
\bibitem{gerhop} M. Gorman, M. el-Hamdi and K. A. Robbins, {\em Hopping
motion of ordered states of cellular flames, }preprint.
\bibitem{gerchaos} M. Gorman, M. el-Hamdi and K. A. Robbins,
{\em Four types of chaotic dynamics in cellular flames, } preprint.
\bibitem{gherrot} M. Gorman, C. F. Hamill, M. el-Hamdi
and K. A. Robbins, {\em Rotating
and modulated rotating states of cellular flames, }preprint.
\bibitem{gs} M. Gorman and H. L. Swinney, {\em Spatial and temporal
characteristics of modulated waves in the circular Couette
system}, J. Fluid Mech. { \bf 117} (1982), 123-142.
\bibitem{go} D. Gottlieb and S. A. Orszag, {\em Numerical Analysis
of Spectral Methods: Theory and Applications, } C.B.M.S. - N.S.F.
Conference Series in Applied Mathematics, SIAM, Philadelphia, (1977).
\bibitem{KC} J. Kevorkian and J. D. Cole,
{\em Perturbation Methods in Applied Mathematics},
Springer-Verlag, New York, (1981).
\bibitem{km} R. Kuske and B. J. Matkowsky, {\em Two dimensional
cellular burner stabilized flames,} Quart. Appl. Math., to appear.
\bibitem{dolit} H. Lofting, {\em The Voyages of Doctor
Dolittle}, J.B. Lippincott, Philadelphia, (1922).
\bibitem{mm} S. B. Margolis and B. J. Matkowsky, {\em Nonlinear
stability and bifurcations in the transition from laminar to
turbulent flame propagation}, Comb. Sci. and Tech.
{\bf 34} (1983), 45-77.
\bibitem{mark} G. H. Markstein, ed., {\em Nonsteady Flame Propagation},
Pergamon Press, Elmsford, NY (1967).
\bibitem{motravel} B. J. Matkowsky and D. O. Olagunju, {\em  Traveling
waves along the front of a pulsating flame, } SIAM J. Appl. Math. {\bf 42}
(1982), 486-501.
\bibitem{mospin} B. J. Matkowsky, and D. O. Olagunju, {\em  Spinning
waves in gaseous combustion,} SIAM J. Appl. Math. {\bf 42} (1982),
1138-1156.
\bibitem{mps} B. J. Matkowsky, L. J. Putnick and G. I. Sivashinsky,
{\em A nonlinear theory of cellular flames,} SIAM J. Appl. Math.
{\bf 38} (1980), 489-504.
\bibitem{msdt} B. J. Matkowsky and G. I. Sivashinsky, {\em An asymptotic
derivation of two models in flame theory associated with the constant
density approximation, }SIAM
J. Appl. Math. {\bf 37} (1979), 686-699.
\bibitem{omq} D. O. Olagunju and B. J. Matkowsky {\em Burner-stabilized
cellular flames,} Quart. J. Appl. Math. {\bf 48} (1990), 645-664.
\bibitem{om} D. O. Olagunju and B. J. Matkowsky {\em Polyhedral
flames,} SIAM J. Appl. Math. {\bf 51} (1991), 73-89.
\bibitem{omc} D. O. Olagunju and B. J. Matkowsky {\em Coupled
complex Ginzburg-Landau type equations
in gaseous combustion,} , Stab. and Appl. Analysis
of Continuous Media {\bf 2} (1992), 31-58.
\bibitem{pj} M.R.E. Proctor and C.A. Jones,
{\it The interaction of two spatially resonant patterns in thermal
convection. Part 1. Exact 1:2 resonance}, J. Fluid Mech. {\bf 188} (1988),
301-335.
\bibitem{coud} M. Rabaud, S. Michalland and Y. Couder,
{\it Dynamical regimes of directional viscous fingering:
 spatiotemporal chaos and wave propagation},
Phys. Rev. Lett. {\bf 64} (1990), 184-187.
\bibitem{r} D. Rand, {\em Dyanmics and symmetry.  Predictions for modulated
waves in rotating fluids, } Arch. Rat. Mech. Anal {\bf 79}
(1982), 1-38.
\bibitem{rr} W.J. Rappel and  H. Riecke,
{\it Parity-breaking in directional solidification: numerics versus amplitude
equations}, Phys. Rev. A {\bf 45} (1992), 846-859.
\bibitem{rp} H. Riecke and H.G. Paap, {\em Parity breaking and Hopf bifurcation
in
axisymmetric Taylor vortex flow}, Phys. Rev. A {\bf 45} (1992), 8605-8610.
\bibitem{sivashinsky} G. I. Sivashinsky, {\em Instabilities, pattern
formation, and turbulence in flames,} Ann. Rev. Fluid Mech.
{\bf 15} (1983), 179-199.
\bibitem{wm92} R.J. Wiener and D.F. McAlister,  {\it Parity-breaking and
solitary
waves in axisymmetric Taylor vortex flow}, Phys. Rev. Lett. {\bf 69} (1992),
2915-2918.
\end{thebibliography}
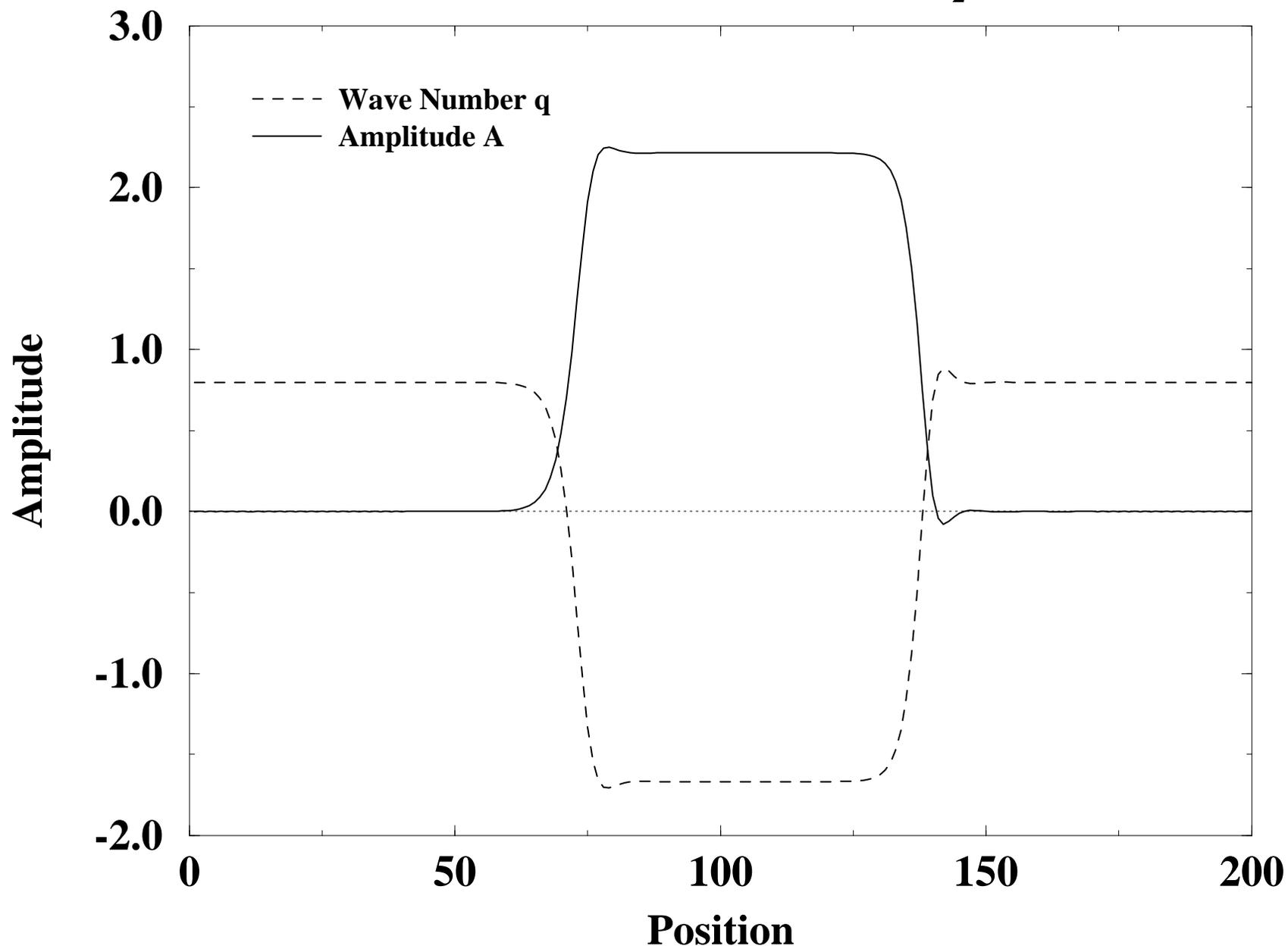
\end{document}